\documentclass[reprint,superscriptaddress,nofootinbib,twocolumn,amsmath,amssymb,aps,prl]{revtex4-2}

\usepackage{amsmath,gensymb,textcomp,bm,dcolumn,eurosym,array,tabu,multirow,nicefrac,color,graphicx,upgreek}
\usepackage[colorlinks, linkcolor=blue, citecolor=blue, urlcolor=blue, breaklinks=true]{hyperref}

\usepackage{mathpazo,times} 

\newcommand{\bra}[1]{\langle #1 \rvert}
\newcommand{\ket}[1]{\lvert #1 \rangle}

\usepackage[subsectionbib]{bibunits}
\usepackage{graphicx}
\usepackage{epstopdf}
\usepackage{epsfig}
\usepackage{dcolumn}
\usepackage{bm,MnSymbol}
\usepackage{units}
\usepackage{psfrag}
\usepackage{color}

\DeclareMathOperator{\sinc}{sinc}

\newcommand{\pare}[1]{\left( #1 \right)}
\newcommand{\ave}[1]{\langle #1 \rangle}
\newcommand{\abs}[1]{\left\vert #1 \right\vert}
\newcommand{\cor}[1]{\left[ #1 \right]}
\newcommand{\llav}[1]{\left\lbrace #1 \right\rbrace}

\begin{document}

\title{Witnessing Entangled Two-Photon Absorption via Quantum Interferometry}

\author{\'Aulide Mart\'inez-Tapia}
\affiliation{Instituto de Ciencias Nucleares, Universidad Nacional Aut\'onoma de M\'exico, Apartado Postal 70-543, 04510 Cd. Mx., M\'exico}

\author{Samuel Corona-Aquino}
\affiliation{Instituto de Ciencias Nucleares, Universidad Nacional Aut\'onoma de M\'exico, Apartado Postal 70-543, 04510 Cd. Mx., M\'exico}


\author{Chenglong You}
\affiliation{Quantum Photonics Laboratory, Department of Physics \& Astronomy, Louisiana State University, Baton Rouge, LA 70803, USA}

\author{Rui-Bo Jin}
\affiliation{Hubei Key Laboratory of Optical Information and Pattern Recognition, Wuhan Institute of Technology, Wuhan 430205, China}

\author{Omar S. Maga\~na-Loaiza}
\affiliation{Quantum Photonics Laboratory, Department of Physics \& Astronomy, Louisiana State University, Baton Rouge, LA 70803, USA}

\author{Shi-Hai Dong}
\email{dongsh2@yahoo.com}
\affiliation{Research Center for Quantum Physics, Huzhou University, Huzhou 313000, China}
\affiliation{Laboratorio de Ciencias de la Informacion Cuantica, CIC, Instituto Politecnico Nacional, UPALM, C.P 07700, CDMX, Mexico}

\author{Alfred B. U'Ren}
\email{alfred.uren@correo.nucleares.unam.mx}
\affiliation{Instituto de Ciencias Nucleares, Universidad Nacional Aut\'onoma de M\'exico, Apartado Postal 70-543, 04510 Cd. Mx., M\'exico}

\author{Roberto de J. Le\'on-Montiel}
\email{roberto.leon@nucleares.unam.mx}
\affiliation{Instituto de Ciencias Nucleares, Universidad Nacional Aut\'onoma de M\'exico, Apartado Postal 70-543, 04510 Cd. Mx., M\'exico}

\begin{abstract}
Recent investigations suggest that the use of non-classical states of light, such as entangled photon pairs, may open new and exciting avenues in experimental two-photon absorption spectroscopy. Despite several experimental studies of entangled two-photon absorption (eTPA), there is still a heated debate on whether eTPA has truly been observed. This interesting debate has arisen, mainly because it has been recently argued that single-photon-loss mechanisms, such as scattering or hot-band absorption may mimic the expected entangled-photon linear absorption behavior. In this work, we focus on transmission measurements of eTPA, and explore three different two-photon quantum interferometers in the context of assessing eTPA. We demonstrate that the so-called N00N-state configuration is the only one amongst those considered insensitive to linear (single-photon) losses. Remarkably, our results show that N00N states may become a potentially powerful tool for quantum spectroscopy, and place them as a strong candidate for the certification of eTPA in an arbitrary sample.    
\end{abstract}
\maketitle


\noindent \textbf{\emph{Introduction.-}}Nonlinear spectroscopy techniques have been shown to constitute a powerful tool for extracting information about the energy dynamics and chemical structure of unknown substances and molecules \cite{ernst_book,mukamel_book,hamm_book,cho_book}. As such, they have played a fundamental role in the development of technologies used in modern society, from process control and manufacturing to pollution monitoring, including also homeland security and healthcare \cite{workman_book}. While in the optical regime these techniques are typically implemented by means of laser light, recent work suggests that the use of non-classical light, such as entangled photon pairs, may offer new and exciting avenues for spectroscopy \cite{dorfman2016,schlawin2017,shi2020,munoz2021,cutipa2022}. 

Remarkably, the time- and frequency-correlations of entangled photon pairs have enabled the observation of non-trivial two-photon absorption phenomena, such as the linear dependence of two-photon absorption rate on the photon flux \cite{javanainen1990,dayan2005,lee2006}. These correlations have been likewise fundamental for predicting fascinating effects, such as two-photon-induced transparency \cite{fei1997,guzman2010}, inducing disallowed atomic transitions \cite{ashok2004}, the manipulation of quantum pathways of matter \cite{roslyak2009,roslyak2009-1,raymer2013,schlawin2016,schlawin2017-1,schlawin2017-2}, and the control of molecular processes \cite{shapiro2011,shapiro_book}. The linear dependence of two-photon absorption as a function of the incident photon flux is particularly relevant, as it implies that one may effectively excite nonlinear phenomena at much lower photon fluxes when compared to classical alternatives  \cite{schlawin2018}. 

During the last decade, entangled two-photon absorption spectroscopy has been identified as a promising tool for extracting the information about the electronic levels that contribute to the two-photon excitation of a molecular sample \cite{saleh,KOJIMA,nphoton,roberto_spectral_shape,oka2010,villabona_calderon_2017,Varnavski2017,oka2018-1,oka2018-2,svozilik2018-1,svozilik2018-2,burdick2018,RobertoTemperatureControlled,Mukamel2020roadmap,Mertenskotter:21}. Moreover, it has been argued that it might provide a new route for probing broadband multi-photon processes with low-power, continuous-wave, single-frequency laser sources \cite{schlawin2018}. Although there is a lively debate on the true quantum enhancement that such a technique might offer to spectroscopy \cite{raymer2021entangled,landes2021quantifying,raymer2021,mikhaylov2022,cushing2022}, the experimental demonstration of its working principle, the so-called entangled two-photon absorption (eTPA), has recently become a topic of keen interest \cite{villabona2020,parzuchowski2021,tabakaev2021,landes2021,samuel2022}. Indeed, it has been argued that the behavior in most of the reported experimental data so far may in fact be due to single-photon-loss phenomena, such as hot-band absorption \cite{mikhaylov2022} or scattering \cite{cushing2022}, which might mimic the sought-after eTPA signals.

Consequently, a large group of physicists, chemists and biologists has devoted considerable effort to developing novel experimental schemes and metrics for certifying true eTPA. Some authors have relied on the linear to quadratic transition in the two-photon absorption rate as a function of the incident photon flux \cite{tabakaev2021}. Others have proposed new metrics based on single and two-photon coincidence measurements, which make use of Hong-Ou-Mandel (HOM)-like interferometers \cite{CIO,samuel2022,cushing2022}. In this work we explore, in the context of eTPA, a number of two-photon quantum interferometry systems, leading to our demonstration that amongst them the so-called N00N-state configuration is the only one insensitive to single-photon losses. Our results show that, in transmission eTPA measurements, this N00N-state configuration is an ideal candidate for the true certification of entangled two-photon absorption. More importantly, and in contrast to other quantum-technology schemes in which N00N states are not typically robust \cite{PhysRevLett.107.113603,Chenglong2021}, our findings show that N00N states could play an important role in quantum spectroscopy.


\begin{figure*}
\centering
\includegraphics[width = 15cm]{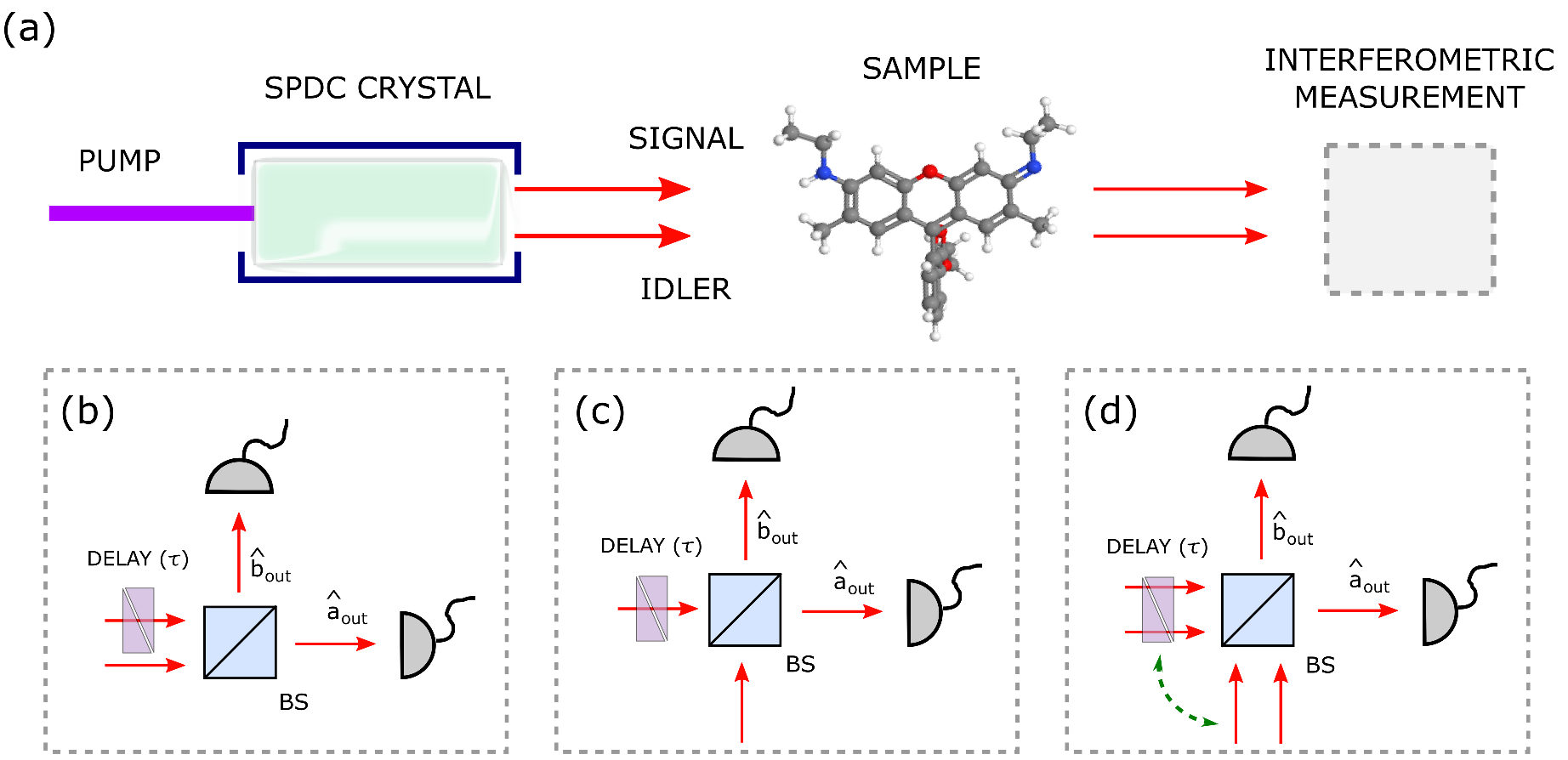}
\caption{Schematic representation of a typical eTPA transmission experiment. (a) Photons pairs, produced by spontaneous parametric down-conversion (SPDC), interact with an arbitrary sample that experiences a two-photon absorption process, along with other single-photon-loss events. The attenuated photon-pair beam is then probed by means of an interferometric coincidence-detection scheme. In previous eTPA experiments \cite{villabona_calderon_2017,samuel2022,cushing2022}, pair losses due to eTPA have been monitored by means of the (b) single-port configuration, where a controllable delay $\tau$ is introduced in one of the photon's path before it impinges on a lossless 50:50 beamsplitter (BS). Other possible interferometric configurations are the (c) two-port configuration, which describes a Hong-Ou-Mandel (HOM) interferometer, and (c) a N00N-state interferometer.}
\label{fig:Fig1}
\end{figure*}

\vspace{5mm}

\noindent \textbf{\emph{Two-photon absorption measurements.-}}In a typical eTPA transmission experiment, pairs of correlated photons, typically produced by spontaneous parametric down-conversion (SPDC), interact with an arbitrary sample [see Fig. \ref{fig:Fig1}(a)]. The incident photon pairs are expected to drive a two-photon transition, so that some pairs can be absorbed and therefore removed from the light beam traversing the sample. In most experiments \cite{villabona_calderon_2017,samuel2022,cushing2022}, the configuration shown in Fig. \ref{fig:Fig1}(b) is used to monitor, through coincidence measurements, the reduced photon-pair flux. Although one might be tempted to think that any pair-loss must be due to eTPA, it has been shown that single-photon-loss mechanisms, such as scattering \cite{cushing2022} and hot-band absorption \cite{mikhaylov2022} can lead to a similar behavior as would be expected for eTPA \cite{samuel2022}. In view of this, it naturally becomes desirable to find a new, single-photon-loss insensitive photon-pair measurement scheme that enables direct certification of eTPA. 

In order to find a good candidate for eTPA certification, we study and compare three distinct two-photon quantum interferometers, namely  single-port and two-port Hong-Ou-Mandel setups, as well as a N00N-state configuration. The three configurations are shown schematically  in Figs. \ref{fig:Fig1}(b)-(d), respectively. Note that the N00N state configuration makes use of a superposition of both photons impinging on input port $a$ and both photons impinging on input port $b$ of the BS. Possible experimental implementations for each of these configurations are discussed in detail in the supplementary materials. In all configurations, two-photon interference takes place via a beam splitter (BS), which for the sake of completeness is assumed for the analysis below to exhibit losses. The input-output transformation of a lossy BS is given by \cite{PhysRevA.57.2134}
\begin{align}
    \hat{a}_{\text{out}} &=  t(\omega)\hat{a}_{\text{in}}(\omega) + r(\omega)\hat{b}_{\text{in}}(\omega) + \hat{F}_a(\omega), \label{eq1}\\
    \hat{b}_{\text{out}} &=  t(\omega)\hat{b}_{\text{in}}(\omega) + r(\omega)\hat{a}_{\text{in}}(\omega) + \hat{F}_b(\omega),\label{eq2}
\end{align}
with $\hat{a}_{\text{in},\text{out}}(\omega)$ and $\hat{b}_{\text{in},\text{out}}(\omega)$ depicting the input and output field modes of the BS, respectively. $r(\omega)$ and $t(\omega)$ are the BS reflection and transmission coefficients, while $\hat{F}_a(\omega)$ and $\hat{F}_b(\omega)$, which commute with the input field operators, are the Langevin noise operators associated with the BS losses \cite{PhysRevA.57.2134}. Note that the Langevin operators lead to reflection and transmission coefficients that do not conserve energy, i.e. $\abs{r(\omega)}^{2} + \abs{t(\omega)}^{2} \leq 1$ \cite{PhysRevA.57.2134}.

We can use Eqs. (\ref{eq1}) and (\ref{eq2}) to monitor the number of photon pairs that impinge on the BS by measuring the photon-coincidence rate at the output of the lossy BS. This can be expressed as 
\begin{equation}
   R = P(1_a, 1_b) = \langle \hat{N}_a\hat{N}_b \rangle, 
\end{equation}
where $\ave{\cdots}$ denotes an expectation value, and the continuum number operators \cite{barnett_book} for the two output ports are given by $\hat{N}_a = \int d\omega\hat{a}^{\dagger}_{\text{out}}(\omega)\hat{a}_{\text{out}}(\omega)$, and $\hat{N}_b = \int d\omega\hat{b}^{\dagger}_{\text{out}}(\omega)\hat{b}_{\text{out}}(\omega)$. 

By considering the simplest case of a lossless 50:50 BS, with $t=\pm ir$, where the $i$ represents the $\pi /2$ phase difference between the transmitted and reflected beams, and $\abs{t}=\abs{r}=1/\sqrt{2}$, we can readily find that the photon-coincidence rate for each of the above configurations (see supplementary materials for details) is given by 
\begin{widetext}
\begin{eqnarray}
        \hspace{-4mm}R_{\pm}(\tau) &=& \frac{1}{4}\int d\nu_s d\nu_i [ \lvert \phi(\nu_s,\nu_i) \rvert^2 + \lvert \phi(\nu_i,\nu_s)\rvert^2  \pm \phi(\nu_s,\nu_i)\phi^*(\nu_i,\nu_s)e^{-i(\nu_i-\nu_s)\tau} \pm \phi^*(\nu_s,\nu_i)\phi(\nu_i,\nu_s)e^{i(\nu_i-\nu_s)\tau}], \label{Eq:pm}\\
        \hspace{-4mm}R_N(\tau) &=& \frac{1}{4} \int d\nu_s d\nu_i [ \lvert \phi(\nu_s,\nu_i)\rvert^2 + \lvert \phi(\nu_i,\nu_s)\rvert^2 + \phi(\nu_s,\nu_i)\phi^*(\nu_i,\nu_s) + \phi^*(\nu_s,\nu_i)\phi(\nu_i,\nu_s) ] \lbrace 1 + \cos[(\nu_s+\nu_i)\tau] \rbrace, \label{Eq:N} 
\end{eqnarray}
\end{widetext}
with $R_{+}(\tau)$, $R_{-}(\tau)$, and $R_{N}\pare{\tau}$ describing the coincidence-count rate for the single-port, two-port and the N00N-state configurations, respectively. The function $\phi(\nu_s,\nu_i)$ represents the signal ($s$) - idler ($i$) joint spectral amplitude. The photon-pair state after interacting with the sample can be written, without loss of generality, as $|\psi\rangle_ = \int  \,d\nu_s \,d\nu_i  \phi(\nu_s,\nu_i) \hat{a}^{\dagger}(\nu_s+\omega_0)\hat{a}^{\dagger}(\nu_i+\omega_0) |0\rangle$. In writing Eqs. (\ref{Eq:pm}) and (\ref{Eq:N}), we have assumed that the photon pairs are frequency degenerate, with a central frequency $\omega_0$. Their frequency deviations from $\omega_0$ are thus given by $\nu_j = \omega_j - \omega_0$ ($j=s,i$). Moreover, note that we have introduced an external delay in one of the input ports of the BS. This delay allows us to  perform a Hong-Ou-Mandel-like measurement on those photon pairs that are not absorbed by the sample. Interestingly, as we will describe below, the coincidence rate as a function of the delay carries important information regarding the nature of the photon-pair losses.

\vspace{5mm}

\noindent \textbf{\emph{ETPA as a two-photon spectral filter.-}}Since its conception, eTPA has been described as a process in which correlated photon pairs satisfying the so-called two-photon resonance condition \cite{Fei-1997,roberto_spectral_shape,RobertoTemperatureControlled} are lost in order to drive a two-photon excitation of the absorbing medium. This means that the sample effectively acts as a \emph{notch filter} that removes specific two-photon resonance frequencies, thus modifying the joint spectral intensity (JSI) of the correlated photons. Mathematically, this transformation can be described by
\begin{equation}
S\pare{\nu_s,\nu_i} = \abs{\phi\pare{\nu_s,\nu_i}}^{2} = \abs{f_{\text{TP}}\pare{\nu_s,\nu_i}\Phi\pare{\nu_s,\nu_i}}^{2},   
\end{equation}
where $\Phi\pare{\nu_s,\nu_i}$ and $\phi\pare{\nu_s,\nu_i}$ describe the joint amplitude of the photons before and after the interaction with the sample, respectively. The two-photon filter can readily be defined as \cite{CIO}
\begin{equation}
f_{\text{TP}}\pare{\nu_s,\nu_i} = 1 - \exp\cor{-\pare{\nu_{s}+\nu_{i}}^2/\pare{2\sigma_{TP}^2}}, \end{equation}
with $\sigma_{TP}$ describing the filter bandwidth. As discussed above, in real experiments, during the light-matter interaction the two-photon beam might experience single-photon losses that remove, independently, signal or idler photons. These losses can be accounted for by writing a single-photon filter of the form
\begin{equation}
f_{s,i}\pare{\nu_s,\nu_i} = 1 - \exp\cor{-\pare{\nu_{s,i}-\nu_{s,i}^{0}}^2/\pare{2\sigma_{s,i}^2}}.
\end{equation}
Here $\nu_{s,i}^{0}$ describe the central frequency deviations of the filter, whereas $\sigma_{s,i}$ represent the single-photon filter bandwidth for the signal and idler modes, respectively.
To understand the effects of the single- and two-photon filters, we assume the most general form for the initial joint spectral intensity (JSI) of the photons (see supplementary materials for details)
\begin{equation}\label{Eq:Phi}
    \begin{split}
\Phi\pare{\nu_s,\nu_i} =& E_{p}\pare{\nu_s,\nu_i}\sinc\cor{L\pare{\eta_s\nu_s+\eta_i\nu_i}/2} \\
&\times \exp\cor{-iL\pare{\eta_s\nu_s+\eta_i\nu_i}/2}.
    \end{split}
\end{equation}
In writing Eq. (\ref{Eq:Phi}) we have used the definition $\sinc\pare{x}=\sin\pare{x}/x$. $E_p\pare{\nu_s,\nu_i} = \exp\cor{-2T_{p}^{2}\pare{\nu_s+\nu_i}^2}$ corresponds to the Gaussian spectral shape of the classical pulsed pump, with temporal duration $T_p$, which pumps the SPDC crystal of length $L$. Finally, $\eta_{s,i} = N_{p}-N_{s,i}$ describes the difference between the inverse group velocity of the pump and the signal and idler photons, respectively. 

\begin{figure}[b!]
    \centering
    \includegraphics[width = 7.5cm]{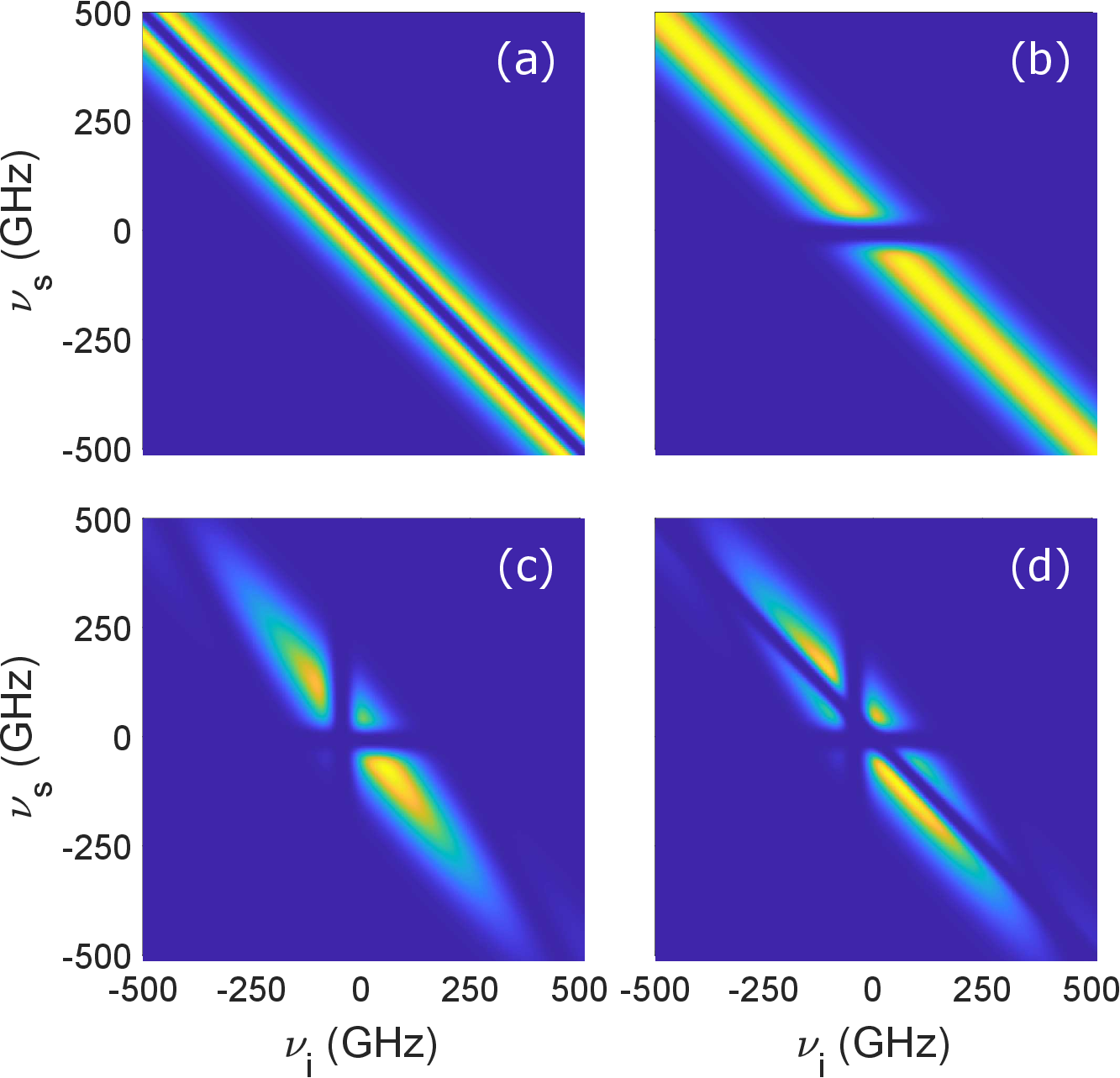}
    \caption{Filtered joint spectral intensities (JSIs). (a) and (b) show symmetric ($\eta_s=\eta_i$) JSIs modified by two-photon and one single-photon filters, respectively. (c) and (d) show asymmetric ($\eta_s\neq\eta_i$) JSIs modified by two single-photon filters in each mode, and a combination of all (two-photon and single-photon) filters, respectively. Note that two-photon losses are depicted by regions along the anti-diagonal [see 2(a)] of the JSI, whereas single-photon losses [2(b) and (c)] correspond to horizontal and vertical lines for the signal and idler modes, respectively. For the symmetric JSIs we use $\eta_s L=\eta_i L = T_p$; whereas for the asymmetric case $\eta_s L= \eta_i L/2 = T_p$. In all cases we set $T_p = 5$ ps, and $\sigma_{\text{TP}}=\sigma_{s}=\sigma_{i}=20$ GHz.}
    \label{Fig2}
\end{figure}

Figure \ref{Fig2} shows some examples of filtered JSIs for (a,b) symmetric (resulting from type 0 or 1 SPDC with $\eta_s=\eta_i$), and (c,d) asymmetric (obtained from type-II SPDC with $\eta_s\neq\eta_i$) initial two-photon states. Note that two-photon losses [Fig. \ref{Fig2}(a)] are characterized by regions along the anti-diagonal of the JSI, whereas single-photon losses [Figs. \ref{Fig2}(b) and (c)] correspond to regions along horizontal or vertical lines for the signal and idler modes, respectively. 

\begin{figure*}
\centering
\includegraphics[width = 16.5cm]{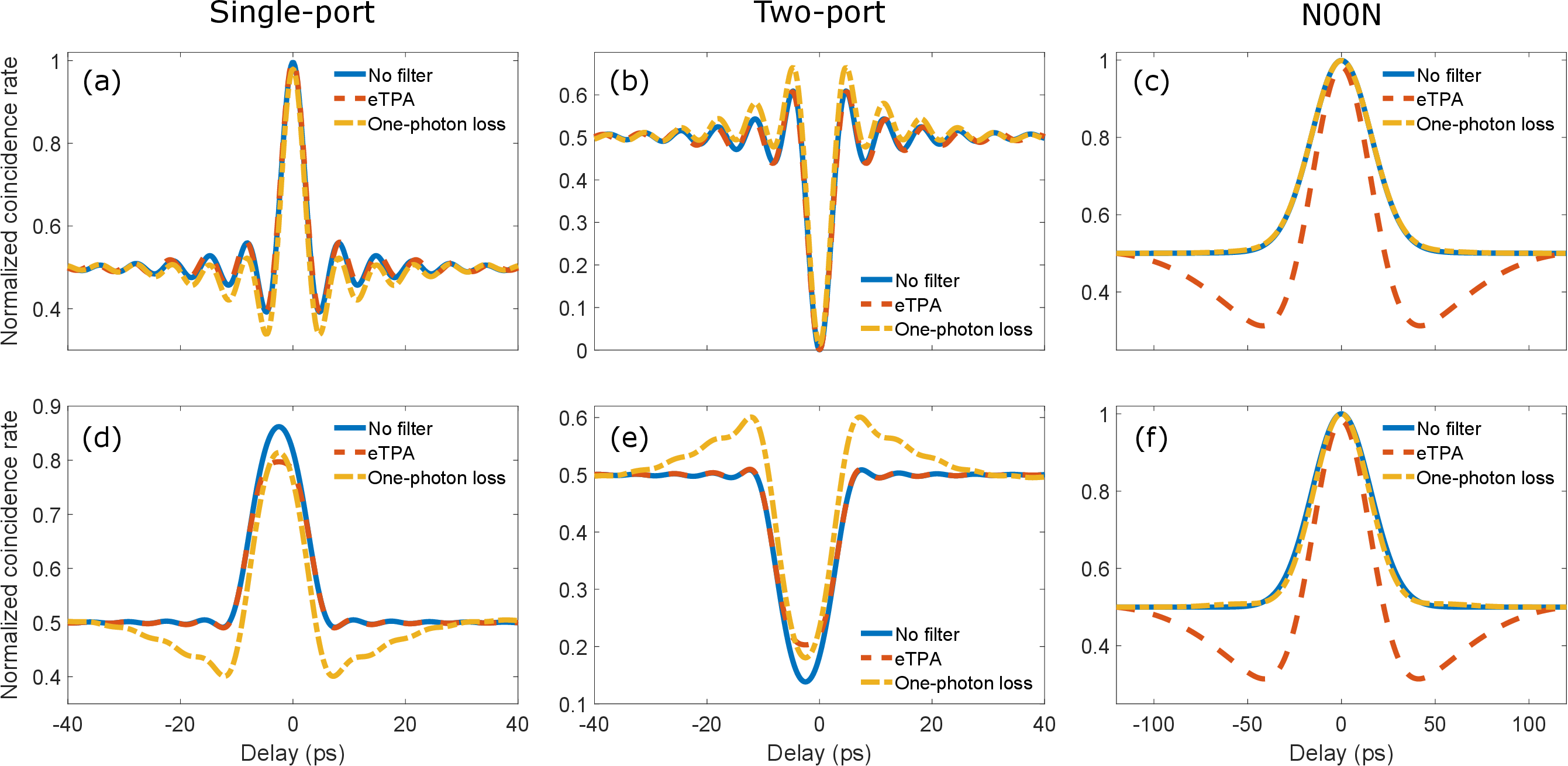}
\caption{Coincidence rates as a function of the delay, $\tau$, for the (a,d) single-port, (b,e) two-port and (d,f) N00N-state configurations. The upper row shows the results for symmetric ($\eta_s=\eta_i$) two-photon states, whereas the bottom row shows the results for asymmetric ($\eta_s\neq\eta_i$) two-photon states. Note that for the N00N-state configuration, the no-filter measurement (blue solid line) and the the one-photon loss (yellow dash-dotted line) curves are essentially fully overlapped, while the eTPA curve (red dashed line) shows an altogether different behavior.  This implies that for symmetric or asymmetric two-photon states, the only scheme amongst those considered that is capable of witnessing pure eTPA is the N00N-state configuration. For the symmetric case, we have set $\eta_s L=\eta_i L = T_p$, while for the asymmetric case we use $\eta_s L= \eta_i L/2 = T_p$. In all cases we set $T_p = 5$ ps, and $\sigma_{\text{TP}}=\sigma_{s}=\sigma_{i}=20$ GHz. }
\label{Fig3}
\end{figure*}


\vspace{5mm}

\noindent \textbf{\emph{Two-photon coincidence rates in the presence of one- and two-photon losses.-}}Having defined the specific form of the two-photon state following the interaction with the sample---where one and two-photon-loss processes may be taking place---we are now ready to evaluate the coincidence rates for each of the previously described interferometric schemes.  Here we consider three cases: i) no filter applied, ii) a two-photon (eTPA) filter applied, and iii)  a linear filter applied to one of the signal and idler photons.   While other possible cases (including a linear filter applied to both photons, as well as the application of both linear and non-linear filters) were considered in our analysis, they do not contribute key, additional physical insights, and were thus omitted from the presented results. 

Figure \ref{Fig3} shows the coincidence rate as a function of the delay $\tau$ for the (a,d) single-port, (b,e) two-port and (c,f) N00N-state configurations. The top row shows the results for initially symmetric ($\eta_s=\eta_i$) two-photon states, whereas the bottom row shows the results for initially asymmetric ($\eta_s\neq\eta_i$) states. For the symmetric case, note that the no-filter (blue solid line) and eTPA (dashed red line) curves are fully overlapped, with the linear losses curve nearly overlapped with the other two,
implying that one would be unable to determine the presence or absence of an eTPA sample from a transmission-based measurement.  
In striking contrast, for the N00N-state configuration, while the no-filter and linear filter curves are fully overlapped, the eTPA curve clearly deviates from the other two.  This means that for symmetric two-photon states, the only scheme that is capable of witnessing eTPA is indeed, the N00N-state configuration. 

Let us now turn to the asymmetric case (bottom row of Fig. \ref{Fig3}). Note that both the two-photon (eTPA) and single photon filter curves exhibit differences with the no-filter curve.  Nevertheless, because the two effects occur together, broadly with a similar behavior,  one may unable to distinguish eTPA from single-photon losses by monitoring changes in the coincidence peak/dip.
It is worth mentioning that in Fig. \ref{Fig3}, the bandwidths of the one- and two-photon loss filters are assumed to be the same. Of course, if one were to suppress single photon losses, e.g. through a reduction of the single photon filter bandwidth, the resulting curve would more closely follow the no-filter curve, thus allowing one to discern the  presence of the eTPA process. This case is ideal, as in practice it is challenging to reliably ensure the absence of linear losses.  Remarkably, in the N00N-state configuration once again the single-photon loss follows the no-filter signal, whereas eTPA clearly shows an altogether  different behavior. It is important to point out that the ``tails'' of the eTPA signal in the N00N-state configuration are two times broader than those of the one-photon-loss signals, shown in Figs. \ref{Fig3}(d) and (e). This implies that in the presence of experimental signal fluctuations, the N00N-state signal would be easier to monitor. Finally, an interesting point to note in Figs. \ref{Fig3}(d) and (e) is that the dip/peak is displaced from $\tau=0$, due to the asymmetry of the two-photon correlation function [Eq. (\ref{Eq:Phi})]. By numerically analyzing the photon-coincidence rate for different values of $\eta_{s}$ and $\eta_{i}$, one can find that the single-port and two-port coincidence rates for asymmetric two-photon states will exhibit a $\pare{\eta_{s}-\eta_{i}}L/2$ temporal delay shift.  

In general, the HOM visibility is unity for a perfectly symmetric JSI (upon the interchange of frequency arguments), while any asymmetry results in a reduction of visibility.  Note that the visibility obtained in the single- and double-port configurations, in all of the cases considered above, may be understood in terms of the symmetry properties of the resulting overall JSI including the effect of any single-photon or two-photon filters.

\vspace{5mm}

\noindent \textbf{\emph{Conclusion.-}}In summary, we have explored three distinct two-photon quantum interferometers that may be used to experimentally certify true eTPA. Remarkably, we have found that the so-called N00N-state configuration is the only one amongst those considered, which is insensitive to linear (single-photon) losses. This unique feature makes such a configuration a strong candidate for effectively certifying the absorption of correlated photon pairs in an arbitrary sample. Given the simplicity of the N00N-state configuration, and in contrast to other schemes for quantum technologies in which N00N states are not typically robust \cite{PhysRevLett.107.113603,Chenglong2021}, we expect them to play an important role in transmission-based  quantum spectroscopy. 

\vspace{5mm}

\noindent \textbf{\emph{Acknowledgements.-}}This work was supported by DGAPA-UNAM under the project UNAM-PAPIIT IN102920. C.Y. and O.S.M-L. acknowledge funding from the U. S. Army Research Office under grant W911NF2210088.
AU thankfully acknowledges support from AFOSR grant FA9550-21-1-0147, from UNAM-PAPIIT grant IN103521, and from CONACyT grant 217559.





\onecolumngrid
\vspace{10mm}

\begin{flushleft}
\Large{\textbf{Supplementary material:\vspace{2mm}\\
\Large{Witnessing Entangled Two-Photon Absorption via Quantum Interferometry}}}
\end{flushleft}
\vspace{0.5cm}

\noindent In this supplementary material we
present: (i) the explicit derivation of the coincidence rate for the three different interferometric configurations described in the main text, (ii) the derivation of the two-photon state that allows us to control the symmetry properties of the photon pair joint spectral intensity, and (iii) some examples of possible experimental schemes for the implementation of the two-photon interferometers discussed in the main text. 

\section*{\large{1. Coincidence rates for different two-photon interference configurations}}

In order to derive the coincidence rate as a function of delay for each interferometric configuration, namely, single-port, two-port, and N00N-state, let us first define the necessary mathematical tools. We start  by writing the input-output transformation of a lossy beam splitter (BS), which is given by
\begin{align}
    \hat{a}_{\text{out}} &=  t(\omega)\hat{a}_{\text{in}}(\omega) + r(\omega)\hat{b}_{\text{in}}(\omega) + \hat{F}_a(\omega), \label{Eq:a_out}\\
    \hat{b}_{\text{out}} &=  t(\omega)\hat{b}_{\text{in}}(\omega) + r(\omega)\hat{a}_{\text{in}}(\omega) + \hat{F}_b(\omega), \label{Eq:b_out}
\end{align}
with $\hat{a}_{\text{in},\text{out}}(\omega)$ and $\hat{b}_{\text{in},\text{out}}(\omega)$ depicting the input and output field modes of the BS, respectively, $r(\omega)$ and $t(\omega)$ are the BS reflection and transmission coefficients, while $\hat{F}_a(\omega)$ and $\hat{F}_b(\omega)$, which commute with the input field operators and satisfy the relations $\hat{F}_a(\omega)\ket{0}=\hat{F}_b(\omega)\ket{0}=0$, are the Langevin noise operators associated with the BS losses  \cite{PhysRevA.57.2134}. 

By making use of the output operators in Eqs. (\ref{Eq:a_out}) and (\ref{Eq:b_out}), we can monitor the number of photon pairs arriving at the BS, by writing the photon coincidence rate at the output of the lossy BS:
\begin{equation}\label{Eq:R_sm}
    R = P(1_a, 1_b) = \langle \hat{N}_a\hat{N}_b \rangle,
\end{equation}
where the continuum number operators \cite{barnett_book} for the two output ports are given by $\hat{N}_a = \int d\omega\hat{a}^{\dagger}_{\text{out}}(\omega)\hat{a}_{\text{out}}(\omega)$, $\hat{N}_b = \int d\omega\hat{b}^{\dagger}_{\text{out}}(\omega)\hat{b}_{\text{out}}(\omega)$, and $\ave{\cdots}$ denotes the expectation value with respect to the initial input two-photon state. 
In what follows, we describe the initial state for each configuration, and evaluate the resulting coincidence rate.

\subsection*{\normalsize{\emph{1.1 Single-port configuration}}}

For this configuration [see Fig. 1(b) of the main manuscript], the input two-photon state reads
\begin{equation}\label{Eq:sp_sm}
    \ket{\Psi_{\text{sp}}} = \int\int d\omega_{s}d\omega_{i}\Phi\pare{\omega_{s},\omega_{i}}e^{i\omega_{s}\tau}\hat{a}_{\text{in}}^{\dagger}\pare{\omega_{s}}\hat{a}_{\text{in}}^{\dagger}\pare{\omega_{i}}\ket{0},
\end{equation}
where $\tau$ represents the temporal delay introduced between the signal and idler photons. We can substitute Eq. (\ref{Eq:sp_sm}) into Eq. (\ref{Eq:R_sm}), and use the transformations given by Eqs. (\ref{Eq:a_out}) and (\ref{Eq:b_out}) to write
\begin{eqnarray}
R_{+}\pare{\tau} &=& \int\int d\omega_{1}d\omega_{2}\bra{\Psi_{\text{sp}}}\hat{a}^{\dagger}_{\text{out}}\pare{\omega_{1}}\hat{b}^{\dagger}_{\text{out}}\pare{\omega_{2}}\hat{a}_{\text{out}}\pare{\omega_{1}}\hat{b}_{\text{out}}\pare{\omega_{2}}\ket{\Psi_{\text{sp}}}, \\
&=& \int\int d\omega_{s}d\omega_{i} \abs{t\pare{\omega_{s}}}^{2}\abs{r\pare{\omega_{i}}}^{2}\abs{\Phi\pare{\omega_{s},\omega_{i}}e^{i\pare{\omega_{s}-\omega_{i}}\tau} + \Phi\pare{\omega_{i},\omega_{s}} }.
\end{eqnarray}
Then, by assuming that the reflection and transmission coefficients are approximately constant over the range of frequencies for which $\abs{\Phi\pare{\omega_{s},\omega_{i}}}$ is significant, we arrive to the expression
\begin{equation}
   R_{+}\pare{\tau} = \abs{t}^{2}\abs{r}^{2} \int\int d\omega_{s}d\omega_{i} \abs{\Phi\pare{\omega_{s},\omega_{i}}e^{i\pare{\omega_{s}-\omega_{i}}\tau} + \Phi\pare{\omega_{i},\omega_{s}} }.  
\end{equation}
Finally, by considering the simplest case of a lossless 50:50 BS, where $t=\pm ir$ and $\abs{t}=\abs{r}=1/\sqrt{2}$, we can use the definition of frequency deviations, i.e. $\nu_j = \omega_j - \omega_0$ ($j=s,i$), to obtain the expression shown in Eq. (4), with the sign $+$, of the main text. 
\vspace{5mm}

\subsection*{\normalsize{\emph{1.2 Two-port configuration}}}

In this configuration [see Fig. 1(c) of the main manuscript], the input two-photon state takes the form
\begin{equation}\label{Eq:tp_sm}
    \ket{\Psi_{\text{tp}}} = \int\int d\omega_{s}d\omega_{i}\Phi\pare{\omega_{s},\omega_{i}}e^{i\omega_{s}\tau}\hat{a}_{\text{in}}^{\dagger}\pare{\omega_{s}}\hat{b}_{\text{in}}^{\dagger}\pare{\omega_{i}}\ket{0}.
\end{equation}    
Note that the creation operator for the idler photon is now replaced by the $\hat{b}_{\text{in}}$ operator. This means that the idler photon now impinges onto the $b$ input-port of the BS. Then, by substituting Eqs. (\ref{Eq:a_out}), (\ref{Eq:b_out}) and (\ref{Eq:tp_sm}) into Eq. (\ref{Eq:R_sm}), we can write the expression
\begin{eqnarray}
R_{-}\pare{\tau} &=& \int\int d\omega_{1}d\omega_{2}\bra{\Psi_{\text{tp}}}\hat{a}^{\dagger}_{\text{out}}\pare{\omega_{1}}\hat{b}^{\dagger}_{\text{out}}\pare{\omega_{2}}\hat{a}_{\text{out}}\pare{\omega_{1}}\hat{b}_{\text{out}}\pare{\omega_{2}}\ket{\Psi_{\text{tp}}}, \\
&=& \int\int d\omega_{s}d\omega_{i} \bigg[ \abs{t\pare{\omega_{s}}}^{2}\abs{t\pare{\omega_{i}}}^{2}\abs{\Phi\pare{\omega_{s},\omega_{i}}}^2 + \abs{r\pare{\omega_{s}}}^{2}\abs{r\pare{\omega_{i}}}^{2}\abs{\Phi\pare{\omega_{i},\omega_{s}}}^2 \nonumber \\
& & \hspace{21mm} + t^{*}\pare{\omega_{s}}t^{*}\pare{\omega_{i}}r\pare{\omega_{s}}r\pare{\omega_{i}}\Phi^{*}\pare{\omega_{s},\omega_{i}}\Phi\pare{\omega_{i},\omega_{s}}e^{-i\pare{\omega_{s}-\omega_{i}}\tau} \nonumber \\
& & \hspace{21mm} + r^{*}\pare{\omega_{s}}r^{*}\pare{\omega_{i}}t\pare{\omega_{s}}t\pare{\omega_{i}}\Phi\pare{\omega_{s},\omega_{i}}\Phi^{*}\pare{\omega_{i},\omega_{s}}e^{i\pare{\omega_{s}-\omega_{i}}\tau}\bigg],
\end{eqnarray}
which upon assuming that the reflection and transmission coefficients are approximately constant over the range of frequencies for which $\abs{\Phi\pare{\omega_{s},\omega_{i}}}$ is significant, it takes the form
\begin{eqnarray}
R_{-}\pare{\tau} &=& \int\int d\omega_{s}d\omega_{i} \bigg[ \abs{t}^{4}\abs{\Phi\pare{\omega_{s},\omega_{i}}}^2 + \abs{r}^{4}\abs{\Phi\pare{\omega_{i},\omega_{s}}}^2 + t^{*^{2}}r^{2}\Phi^{*}\pare{\omega_{s},\omega_{i}}\Phi\pare{\omega_{i},\omega_{s}}e^{-i\pare{\omega_{s}-\omega_{i}}\tau} \nonumber \\
& & \hspace{21mm} + r^{*^{2}}t^{2}\Phi\pare{\omega_{s},\omega_{i}}\Phi^{*}\pare{\omega_{i},\omega_{s}}e^{i\pare{\omega_{s}-\omega_{i}}\tau}\bigg].
\end{eqnarray}
Finally, by considering the case of the 50:50 BS, with $t=\pm ir$ and $\abs{t}=\abs{r}=1/\sqrt{2}$, we use the frequency deviations, i.e. $\nu_j = \omega_j - \omega_0$ ($j=s,i$), to obtain the expression shown in Eq. (4), with the sign $-$, of the main text. 
\vspace{1mm}

\subsection*{\normalsize{\emph{1.3 N00N-state configuration}}}

The last configuration that we analyze in this work corresponds to a superposition of both photons entering through input port $a$ and both photons entering through input port $b$ of the BS, thus the name \emph{N00N-state} configuration. In this scenario, the two-photon input state takes the form
\begin{equation}\label{Eq:N_sm}
    \ket{\Psi_{\text{N}}} = \frac{1}{\sqrt{2}} \int\int d\omega_{s}d\omega_{i}\Phi\pare{\omega_{s},\omega_{i}}\cor{e^{i\omega_{s}\tau} e^{i\omega_{i}\tau}\hat{a}_{\text{in}}^{\dagger}\pare{\omega_{s}}\hat{a}_{\text{in}}^{\dagger}\pare{\omega_{i}} + \hat{b}_{\text{in}}^{\dagger}\pare{\omega_{s}}\hat{b}_{\text{in}}^{\dagger}\pare{\omega_{i}}}\ket{0}.
\end{equation} 
Note that in this case, we introduce an external delay $\tau$ in one of the possible paths that the photons follow before impinging onto the BS. By substituting Eqs. (\ref{Eq:a_out}), (\ref{Eq:b_out}) and (\ref{Eq:N_sm}) into Eq. (\ref{Eq:R_sm}), we can thus write
\begin{eqnarray}
R_{N}\pare{\tau} &=& \int\int d\omega_{1}d\omega_{2}\bra{\Psi_{\text{N}}}\hat{a}^{\dagger}_{\text{out}}\pare{\omega_{1}}\hat{b}^{\dagger}_{\text{out}}\pare{\omega_{2}}\hat{a}_{\text{out}}\pare{\omega_{1}}\hat{b}_{\text{out}}\pare{\omega_{2}}\ket{\Psi_{\text{N}}}, \\
&=& \frac{1}{2}\int\int d\omega_{s}d\omega_{i} \bigg\{ \bigg[t\pare{\omega_{s}}r\pare{\omega_{i}}e^{i\pare{\omega_{s}+\omega_{i}}\tau} + t\pare{\omega_{i}}r\pare{\omega_{s}}\bigg]\bigg[\Phi\pare{\omega_{s},\omega_{i}} + \Phi\pare{\omega_{i},\omega_{s}}\bigg]\bigg\} \nonumber \\
& & \hspace{5mm} \times \bigg\{\bigg[t^{*}\pare{\omega_{s}}r^{*}\pare{\omega_{i}}e^{-i\pare{\omega_{s}+\omega_{i}}\tau} + t^{*}\pare{\omega_{i}}r^{*}\pare{\omega_{s}}\bigg]\bigg[\Phi^{*}\pare{\omega_{s},\omega_{i}} + \Phi^{*}\pare{\omega_{i},\omega_{s}}\bigg]\bigg\}. \label{Eq:RN}
\end{eqnarray}
By assuming that the reflection and transmission coefficients are approximately constant over the range of frequencies for which $\abs{\Phi\pare{\omega_{s},\omega_{i}}}$ is significant, Eq. (\ref{Eq:RN}) takes the form
\begin{eqnarray}
R_{N}\pare{\tau} &=& \frac{\abs{t}^{2}\abs{r}^{2}}{2}\int\int d\omega_{s}d\omega_{i} \bigg\{ \bigg[ e^{i\pare{\omega_{s}+\omega_{i}}\tau} + 1 \bigg]\bigg[\Phi\pare{\omega_{s},\omega_{i}} + \Phi\pare{\omega_{i},\omega_{s}}\bigg]\bigg\} \nonumber \\
& & \hspace{21mm} \times \bigg\{ \bigg[ e^{-i\pare{\omega_{s}+\omega_{i}}\tau} + 1 \bigg]\bigg[\Phi^{*}\pare{\omega_{s},\omega_{i}} + \Phi^{*}\pare{\omega_{i},\omega_{s}}\bigg]\bigg\}.
\end{eqnarray}
Finally, we consider a lossless 50:50 BS, with $t=\pm ir$ and $\abs{t}=\abs{r}=1/\sqrt{2}$, and use the frequency deviations to find the expression shown in Eq. (5) of the main text.

\section*{\large{2. General two-photon state from an SPDC source }}

To derive the general form of the two-photon state used in the main text, we start by writing the state of a down-converted field at the output plane of the nonlinear, SPDC crystal as \cite{perina}
\begin{equation}
\ket{\Psi} = \int\int d\nu_s d\nu_i \Phi\pare{\nu_s,\nu_i} \ket{\nu_s+\omega_{s}^{0}}\ket{\nu_{i}+\omega_{i}^{0}},
\end{equation}
where the function describing the correlations between down-converted photons (so-called mode function) is given by
\begin{equation} \label{Eq:Phi}
    \Phi\pare{\nu_s,\nu_i} = E_{p}\pare{\nu_s,\nu_i}\sinc\llav{L\cor{k_{p}\pare{\nu_p} - k_{s}\pare{\nu_s} - k_{i}\pare{\nu_i}}/2}\exp\llav{-iL\cor{k_{p}\pare{\nu_p} - k_{s}\pare{\nu_s} - k_{i}\pare{\nu_i}}}. 
\end{equation}
Here we use the definition $\sinc\pare{x}=\sin\pare{x}/x$. $E_{p}\pare{\nu_s,\nu_i}$ is the spectral shape of the pump beam, $L$ is the length of the nonlinear crystal, and $k_{j}$ with $j=p,s,i$ are the wavenumbers of the pump, the signal and the idler photons, respectively. Moreover, the frequency deviations for the three fields are
\begin{eqnarray}
    \nu_{s} &=& \omega_s - \omega_{s}^{0}, \\
    \nu_{i} &=& \omega_i - \omega_{i}^{0}, \\
    \nu_{p} &=& \nu_s + \nu_{i},  \label{Eq:nup}
\end{eqnarray}
with $\omega_{s}^{0}$ and $\omega_{i}^{0}$ being the central frequency of the signal and idler photons, respectively. For the wavenumbers, we expand them in Taylor series and up to first order they read \cite{torres}
\begin{eqnarray}
k_{p}\pare{\nu_p} &\simeq& k_{p}^{0} + N_{p}\nu_{p},  \label{Eq:kp} \\
k_{s}\pare{\nu_s} &\simeq& k_{s}^{0} + N_{s}\nu_{s},  \label{Eq:ks} \\ 
k_{i}\pare{\nu_i} &\simeq& k_{i}^{0} + N_{i}\nu_{i}, \label{Eq:ki}
\end{eqnarray}
where $N_j$ stands for the inverse group velocity of each field---namely, pump, signal and idler photons--and the central wavenumbers, defined as $k_{j}^{0}=\omega_{j}^{0} /c$, satisfy the momentum-conservation relation: $k_{p}^{0} = k_{s}^{0}+k_{i}^{0}$. 

Finally, by substituting Eqs. (\ref{Eq:nup})--(\ref{Eq:ki}) into Eq. (\ref{Eq:Phi}) we can write
\begin{equation} \label{Eq:final_Phi}
    \Phi\pare{\nu_s,\nu_i} = E_{p}\pare{\nu_s+\nu_i}\sinc\cor{L\pare{\eta_s\nu_s+\eta_i\nu_i}/2}\exp\cor{-iL\pare{\eta_s\nu_s+\eta_i\nu_i}/2},
\end{equation}
with $\eta_{s,i} = N_{p}-N_{s,i}$ describing the difference between the inverse group velocity of the pump and the signal and idler photons, respectively. Note that this result corresponds to Eq. (9) of the main text. 

\vspace{5mm}
\section*{\large{3. Experimental implementation of two-photon interferometric measurements }}

In this section, we discuss some examples of possible implementations for the two-photon interferometric configurations discussed in the main text. Figure S1 shows the schematic representation for each configuration. In all cases, for the sake of simplicity, we have assumed type-II SPDC, where down-converted photons exhibit orthogonal (vertical-horizontal) polarizations.

For the single-port configuration [Fig. S1(a)], we start by creating the signal and idler modes with a polarizing beamsplitter. A temporal delay, together with a polarization rotation (by means of a half-wave plate) in one of the arms is introduced. After these transformations are applied, both photons are then injected into a second beamsplitter. This creates a two-photon output state in which one of the photons is temoporally delayed with respect to the other, thus effectively implementing the single-port configuration measurement, depicted in Fig. 1(b) of the main text.

As for the two-port configuration [Fig. S1(b)], we follow the same initial procedure as in the previous case to introduce the external time-delay in one of the photons.  A half-wave plate rotates the polarization in one of the arms so that two co-polarized photons impinge on a BS through its two input ports.

Finally, for the N00N-state configuration [Fig. S1(c)], we initially implement a Hong-Ou-Mandel interferometer, in which the signal and idler photons exit the first, non-polarizing beam splitter in a balanced superposition of both photons exiting (together) in either output port.
Such a balanced superposition is ensured by making the two input path lengths equal. We then introduce an external delay in one of the output modes, before injecting them in the second beam splitter that performs the N00N-state configuration measurement.

\renewcommand{\thefigure}{S1}

\begin{figure}[t!]
    \centering
    \includegraphics[width = 11.5cm]{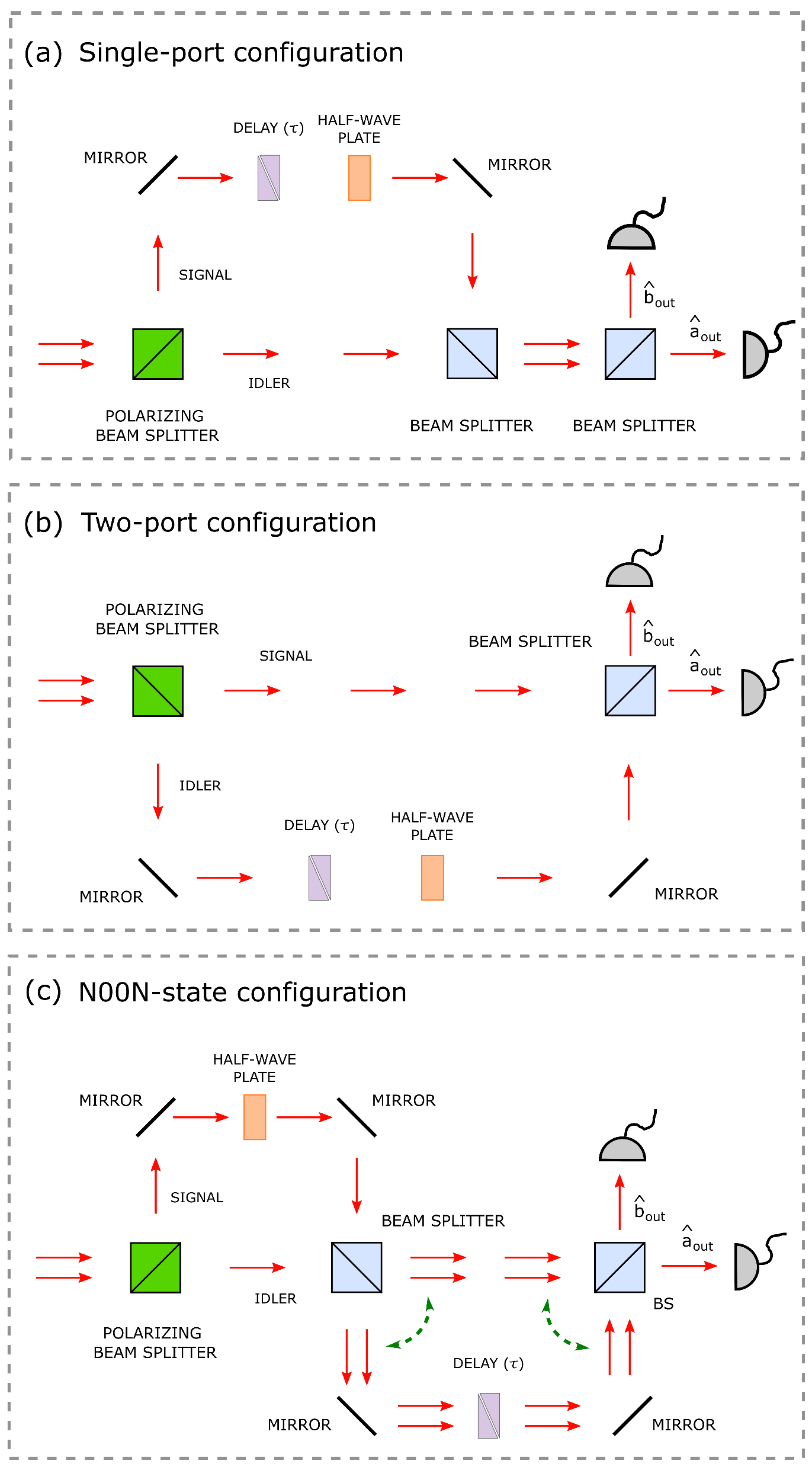}
    \caption{Schematic representation of possible experimental implementations of the two-photon interferometric configurations discussed in the main text, namely (a) single-port, (b) two-port, and (c) N00N-state configurations. For the sake of simplicity, we have assumed that photon pairs are produced by means of type-II SPDC, where down-converted photons exhibit orthogonal (vertical-horizontal) polarizations. This allows us to create the \emph{signal} and \emph{idler} modes of the attenuated (or filtered) beam by making use of a polarizing beamsplitter. }
    \label{Fig2}
\end{figure}

\newpage 

\bibliography{Bib_ETPA}

\end{document}